\documentclass[11pt,a4paper]{iopart}
\usepackage{varioref}
\usepackage[dvips]{graphicx}
\usepackage[latin1]{inputenc}
\usepackage{iopams}  
\begin{document}

\title[]{Recent results on strangeness production from NA49}

\author{Michael Mitrovski for the NA49 Collaboration\footnote[1] {Presented at Strangeness in Quark Matter 2009, Buzios, Rio de Janeiro, Brazil}}

\address{Frankfurt Institute for Advanced Studies (FIAS), Frankfurt, Germany and}
\address{J. W. Goethe-Universit\"at, Frankfurt, Germany}
\ead{Michael.Mitrovski@cern.ch}

\begin{abstract}
We present a summary of measurements of strange particle production performed by the experiment NA49 in inelastic p+p interactions, as well as semi-central C+C and Si+Si, central Pb+Pb, and minimum bias Pb+Pb collisions in the energy range $\sqrt{s_{NN}}$ = 6.3 - 17.3 GeV. New results on $\pi^{-}$, $K^{+}$ and $K^{-}$ production in minimum bias Pb+Pb collisions at $\sqrt{s_{NN}}$ = 8.7 and 17.3 are shown. Furthermore the strangeness enhancement factor at $\sqrt{s_{NN}}$ = 17.3 GeV is presented and compared to the results from NA57 and STAR. Energy dependence of strange particle yields normalized to pion yields is presented. New data on $\left<K^{*}(892)^{0}\right>$ production are shown at $\sqrt{s_{NN}}$ = 17.3 GeV. Furthermore we present the energy dependence of $K/\pi$ and $K/p$ fluctuations. The data are compared with model predictions.

\end{abstract}


\section{Introduction}

Relativistic heavy ion collisions offer a possibility to study nuclear matter under extreme collisions, at high temperatures and densities, in the laboratory. It is expected that at high temperature and baryon density, nuclear matter melts into a state of free quarks and gluons, known as the quark gluon plasma (QGP)~\cite{Bass:1998vz,Wang:1996yf,Harris:1996zx}.

The study of strange hadron production is a good diagnostic tool to investigate the properties of matter created in heavy ion collisions. The measured final distributions of strange hadrons may give insight into strangeness production in the hot and dense stage of the reaction.

Almost 25 years ago the authors in~\cite{Rafelski:1982pu,Koch:1986ud} suggested that strange particle production should be enhanced in collisions with QGP production in comparison to those without QGP and could thus serve as a unique signature of the quark gluon plasma. This enhancement can be quantified by the ratio of yields in nucleus+nucleus to p+p collisions where the system is very small and the QGP creation is not expected.

\begin{figure}
\begin{center}
\includegraphics[scale=0.7]{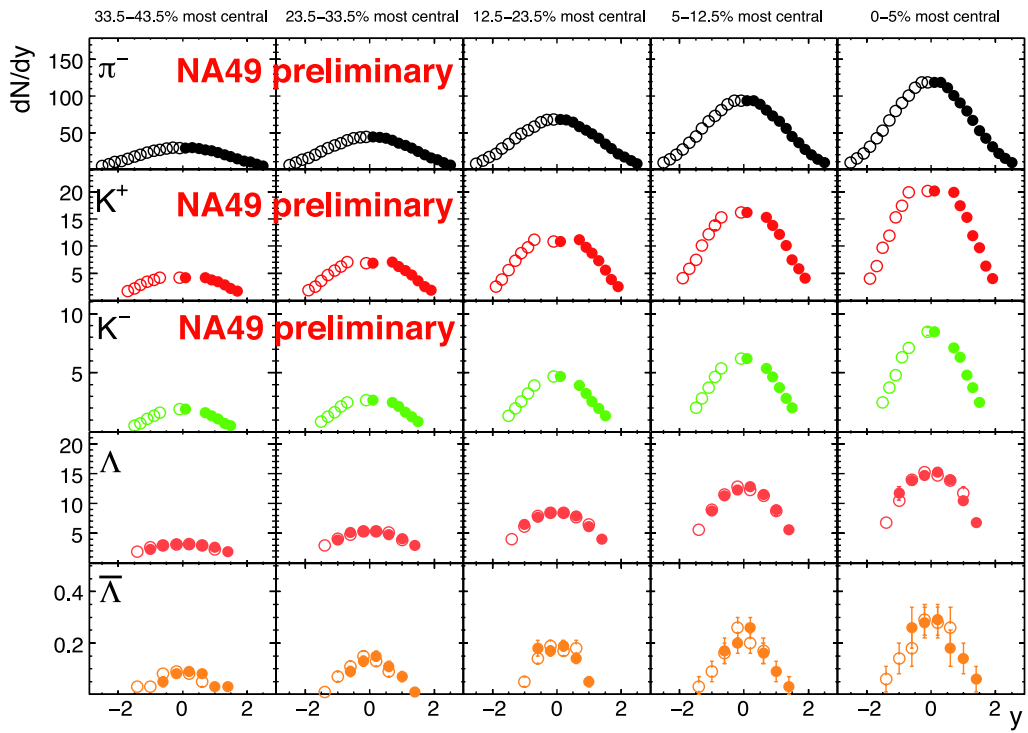}
\caption{\label{fig:fig1} Rapidity distributions for $\pi^{-}$, $K^{+}$, $K^{-}$, $\Lambda$ and $\bar{\Lambda}$~\cite{Anticic:2009ie} measured for different centralities in Pb+Pb collisions at $\sqrt{s_{NN}}$ = 8.7 GeV. The closed symbols indicate measured points, whereas the open symbols are reflected with respect to midrapidity.}
\end{center}
\end{figure}

\section{The NA49 experiment}

The NA49 detector~\cite{Afanasev:1999iu} is a large acceptance hadron spectrometer at the CERN SPS, featuring four large volume TPCs as tracking detectors. Two of them are positioned inside two superconducting dipole magnets. The ionization energy loss (dE/dx) measurements are used for particle identification in the TPCs with a typical resolution of 4\% at forward rapidities. The dE/dx resolution allows in the laboratory frame of the Main-TPCs hadron identification for p $>$ 4 GeV/c with typical K/$\pi$ and K/p separation power of $\approx~$1.5$~\sigma$. Furthermore the two time-of-flight walls (resolution $\sigma_{TOF}$~=~60~ps) improve the separation power in selected regions around midrapidity. The centrality of a given reaction is  determined on the basis of the energy of the projectile spectator nucleons measured by the forward calorimeter. 

\section{Results}

The NA49 experiment has collected data on central Pb+Pb collisions at 5 beam energies in the range $\sqrt{s_{NN}}$ = 6.3 - 17.3 GeV and minimum bias Pb+Pb collisions at $\sqrt{s_{NN}}$  = 8.7 GeV and 17.3 GeV. Also data on C+C and Si+Si collisions as well as p+p interactions where taken at $\sqrt{s_{NN}}$ = 17.3 GeV. 

\subsection{System-size dependence of particle yields}

\begin{figure}
\begin{center}
\includegraphics[scale=0.7]{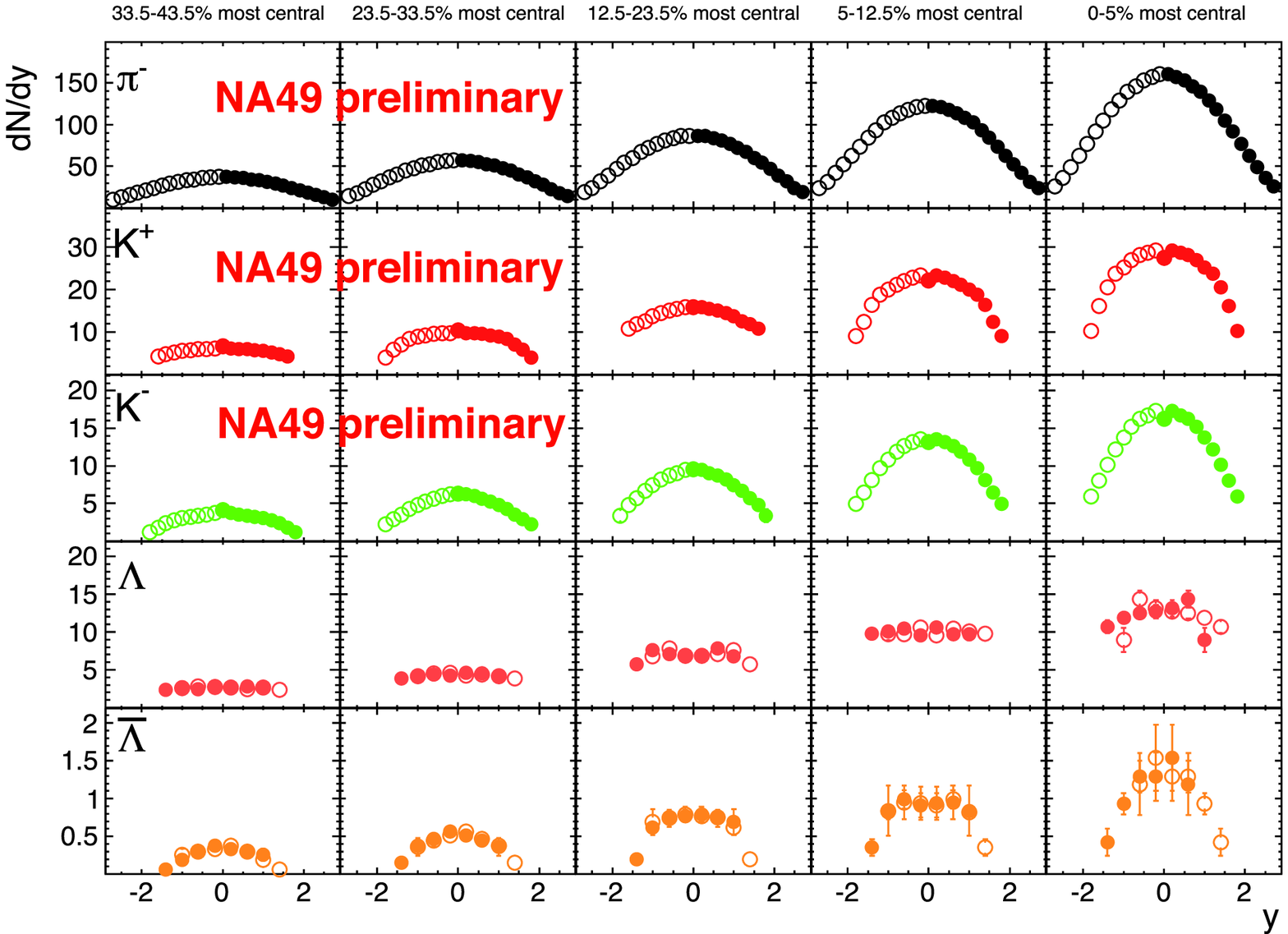} 
\caption{\label{fig:fig2} Rapidity distributions for $\pi^{-}$, $K^{+}$, $K^{-}$, $\Lambda$ and $\bar{\Lambda}$~\cite{Anticic:2009ie} measured for different centralities in Pb+Pb collisions at $\sqrt{s_{NN}}$ = 17.3 GeV. The closed symbols indicate measured points, whereas the open symbols are reflected with respect to midrapidity.}
\end{center}
\end{figure}

The NA49 experiment features a large acceptance in the forward hemisphere allowing measurements of rapidity spectra from midrapidity up to almost beam rapidity. Figures~\ref{fig:fig1} and~\ref{fig:fig2} show rapidity spectra for $\pi^{-}$, $K^{+}$, $K^{-}$, $\Lambda$ and $\bar{\Lambda}$~\cite{Anticic:2009ie} for different centralities in Pb+Pb collisions at $\sqrt{s_{NN}}$ = 8.7 and 17.3 GeV, where the closed points represent the measured data and the open symbols represent the reflected points at midrapidity. 

The total multiplicity is extracted by fitting a (double-) gaussian distribution to the rapidity spectra. It is visible that the width of the pion rapidity-spectra changes with centrality. The distribution become broader when moving to peripheral collisions which might indicate that the influence of resonance decays is much more pronounced for peripheral collisions. In comparison the width of the charged kaons does not change with the centrality of the collision. 

\begin{figure}
\begin{center}
\includegraphics[scale=0.6]{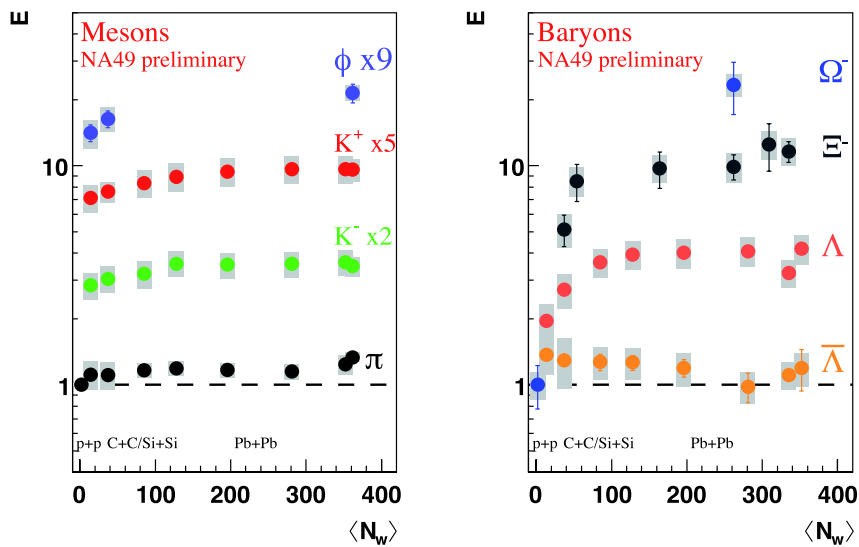}
\caption{\label{fig:fig3}  Strangeness enhancement E at midrapidity ($\mid$y$\mid$ $\leq$ 0.5) for mesons (left) and baryons (right) as a function of $\left<N_{W}\right>$ at $\sqrt{s_{NN}}$ = 17.3 GeV~\cite{Alt:2005zq,Alt:2004wc,Afanasiev:2002mx,Alt:2008iv,Alt:2008qm,Alt:2004kq}. The grey boxes represent systematic errors.}
\end{center}
\end{figure}

Figure~\ref{fig:fig3} shows the strangeness enhancement factor $E$ at midrapidity ($\mid$y$\mid$ $\leq$ 0.5)  for mesons (left) and baryons (right) as a function of the number of wounded nucleons $\left<N_{W}\right>$, calculated within the Glauber model~\cite{Glauber:1970jm}, at $\sqrt{s_{NN}}$ = 17.3 GeV~\cite{Alt:2005zq,Alt:2004wc,Afanasiev:2002mx,Alt:2008iv,Alt:2008qm,Alt:2004kq}. The strangeness enhancement $E$ is defined as the yield per wounded nucleon in Pb+Pb collisions normalized to the corresponding yield in p+p interactions: 

\begin{equation}
\label{eq:equation1}
E = \left( \frac{Yield}{\left<N_{W}\right>} \right)_{A+A}  / \left( \frac{Yield}{2} \right)_{p+p}
\end{equation}

An enhancement is visible for mesons and baryons. For mesons a clear hierarchy is visible from particles without strange quarks ($\pi$) to particles with two strange quarks ($\phi$) when going from p+p to Pb+Pb collisions. A similar behavior is visible for baryons, except the $\bar{\Lambda}$ which does not show any enhancement. Furthermore the enhancement factor increases fast for $N_{W}$ $<$ 60 and then approximately saturates. Various approaches have been used to explain the centrality/system size dependence of (non-)strange particle production. One approach is the Core-Corona model~\cite{Werner:2007bf,Aichelin:2008mi,Becattini:2008yn,Becattini:2008ya, Hohne:2005ks}. It is assumed that the core of a heavy ion collisions is fully thermalized and strangeness is produced according to equilibrium hadron gas. The core is surrounded by a corona which can be modeled as a superposition of elementary p+p reactions. Another approach is using a hydrodynamic model with a transport model as afterburner~\cite{Petersen:2009zi}. One finds that strangeness equilibration is only reached in Pb+Pb collisions. The equilibrium is also reached in peripheral Pb+Pb collisions, whereas for C+C and Si+Si interactions a better description is obtained with a binary scattering transport approach like UrQMD~\cite{Petersen:2009zi,Petersen:2008kb} which does not assume equilibration. Both approaches provide a good description of the data. 

\begin{figure}
\begin{center}
\includegraphics[scale=0.32]{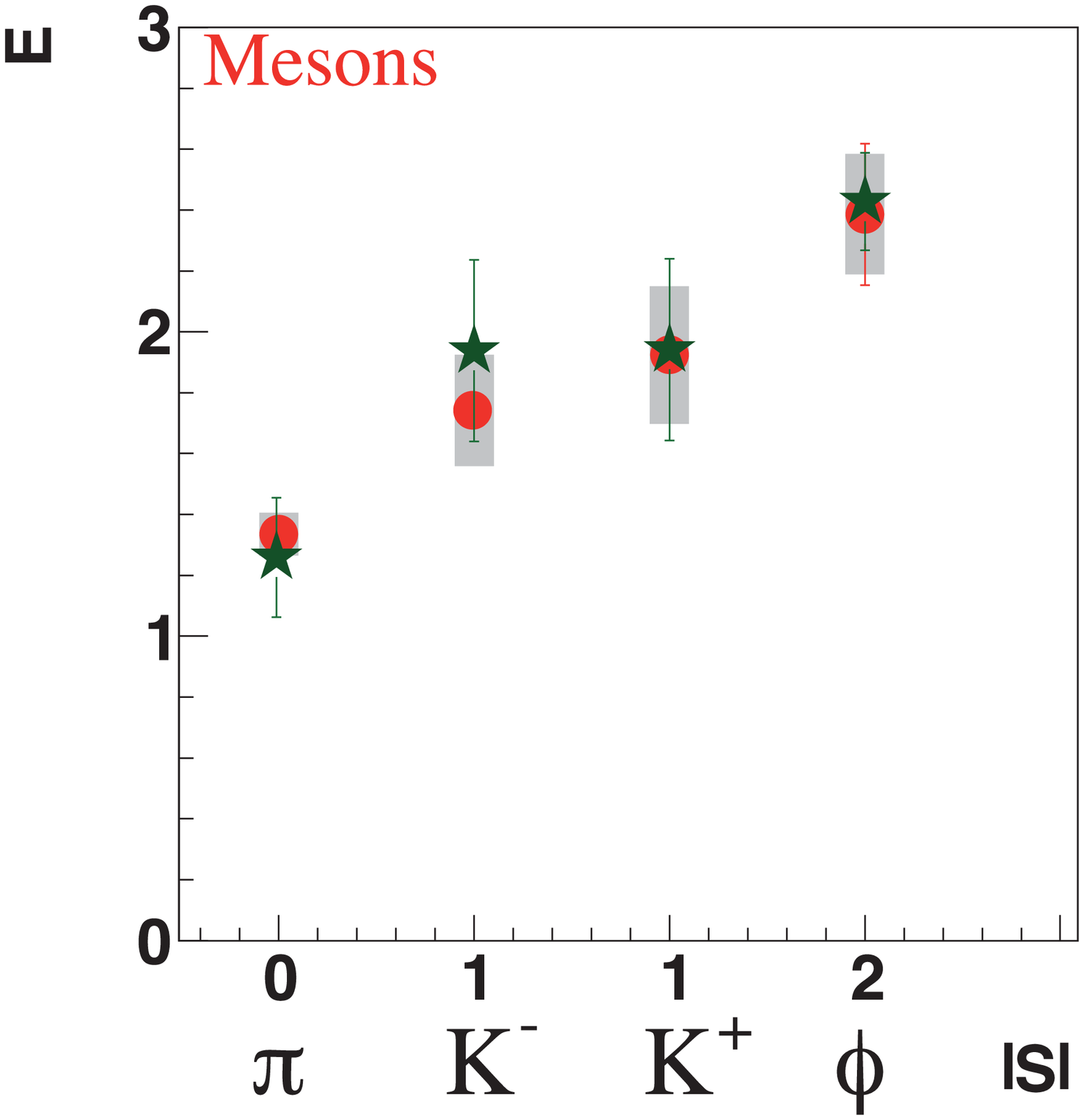}
\includegraphics[scale=0.32]{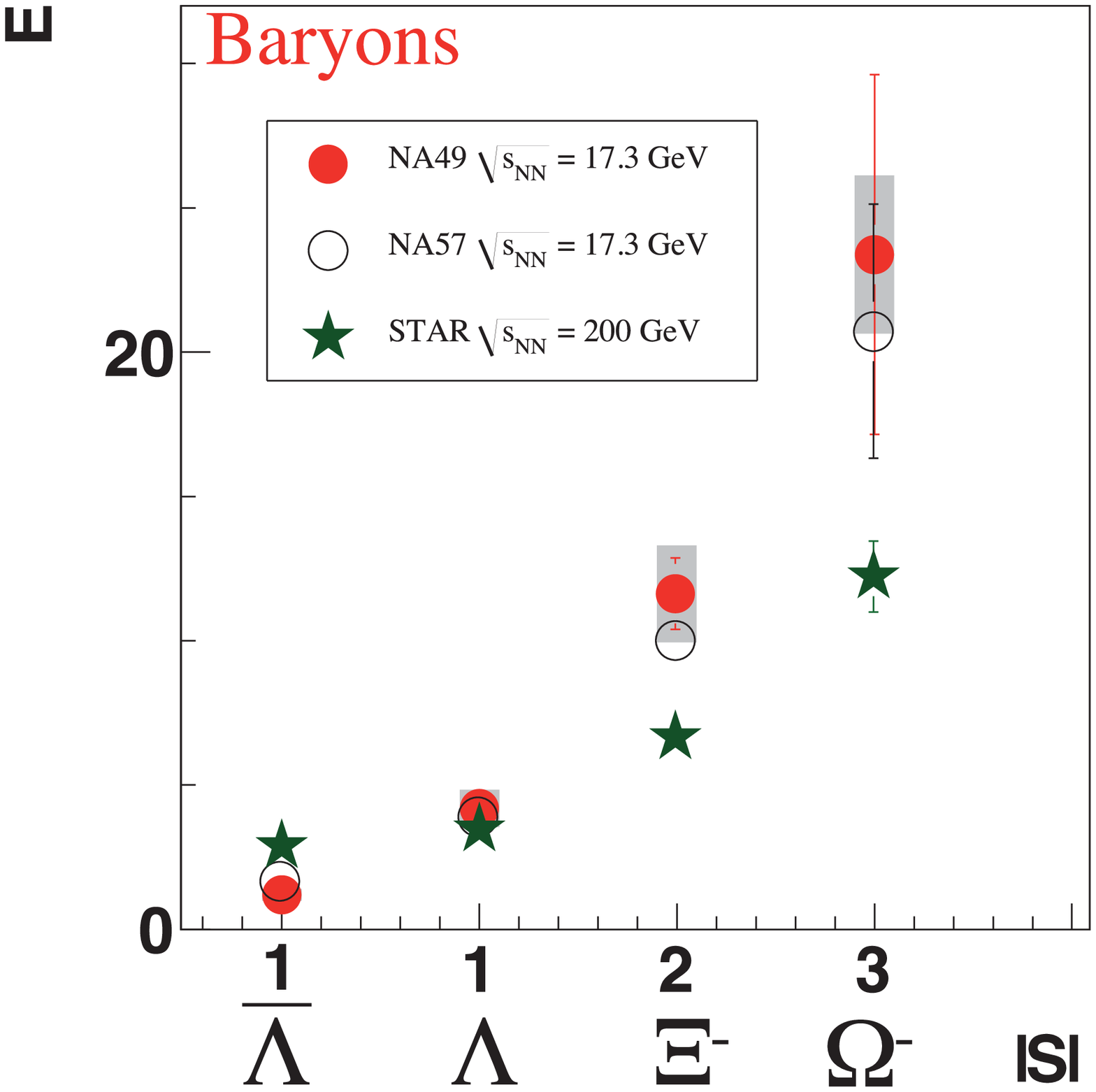}
\caption{\label{fig:fig4}  Enhancement factor E versus strangeness content at midrapidity ($\mid$y$\mid$ $\leq$ 0.5) for mesons (left) and baryons (right) in central Pb+Pb/Au+Au collisions from the NA49~\cite{Alt:2005zq,Alt:2004wc,Afanasiev:2002mx,Alt:2008iv,Alt:2008qm,Alt:2004kq} (closed circles) and NA57~\cite{Antinori:2006ij} (open circles) Collaborations at $\sqrt{s_{NN}}$ = 17.3 GeV and the STAR~\cite{:2008ez,Abelev:2008zk,Abelev:2007xp} (stars) Collaboration at $\sqrt{s_{NN}}$ = 200 GeV (Note NA57 uses p+Be instead of p+p as baseline). The grey boxes represent the systematic error.}
\end{center}
\end{figure}

Figure~\ref{fig:fig4} shows the enhancement factor versus the strangeness content at midrapidity ($\mid$y$\mid$ $\leq$ 0.5) for mesons (left) and baryons (right) in central Pb+Pb/Au+Au collisions from the NA49~\cite{Alt:2005zq,Alt:2004wc,Afanasiev:2002mx,Alt:2008iv,Alt:2008qm,Alt:2004kq} (closed circles) and NA57~\cite{Antinori:2006ij} (open circles) experiments at $\sqrt{s_{NN}}$ = 17.3 GeV and the STAR~\cite{:2008ez,Abelev:2008zk,Abelev:2007xp} (stars) experiment at $\sqrt{s_{NN}}$ = 200 GeV. Clearly, for mesons no difference in the enhancement factors between top SPS and top RHIC energies is observed. For baryons a different behavior is visible. Compared to top RHIC energies the net-baryon density at SPS energies is higher and therefore the $\bar{\Lambda}$s are getting absorbed during the late stage of the evolution~\cite{Petersen:2009zi,Weber:2002qb}. For other strange baryons a hierarchy with the strangeness content is visible but a higher enhancement is observed at top SPS compared to top RHIC energies. The strangeness enhancement factor increases further when going from the SPS to AGS energies. This trend can be understood due to an increasing suppression of strangeness production in p+p interactions with decreasing energy because of energy conservation~\cite{Gazdzicki:2000ht}. Thus the old idea of strangeness enhancement as QGP signal~\cite{Rafelski:1982pu,Koch:1986ud} does not hold at low energies.

\subsection{Energy dependence of particle yields}

\begin{figure}
\begin{center}
\includegraphics[scale=0.42]{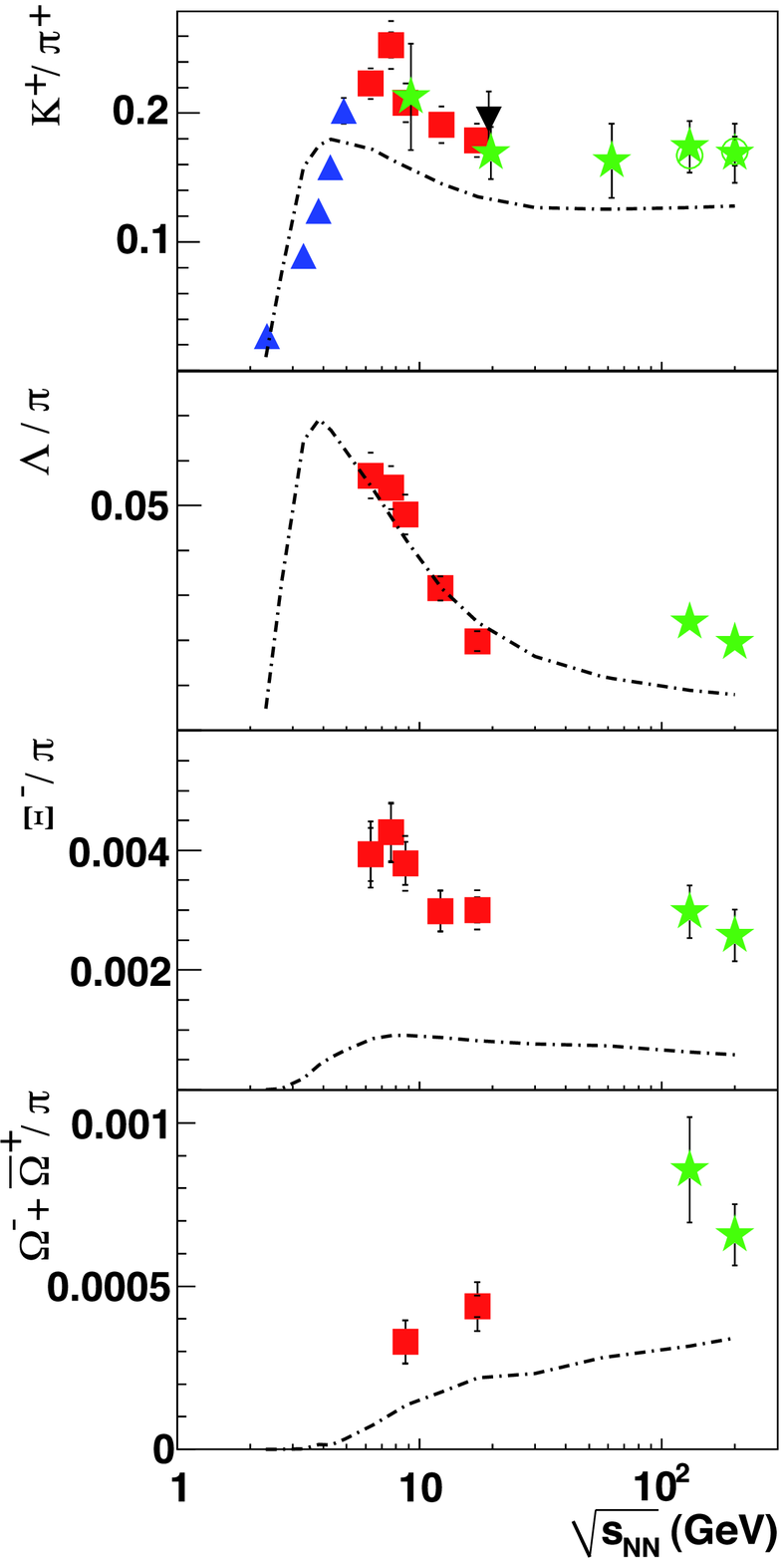}
\includegraphics[scale=0.42]{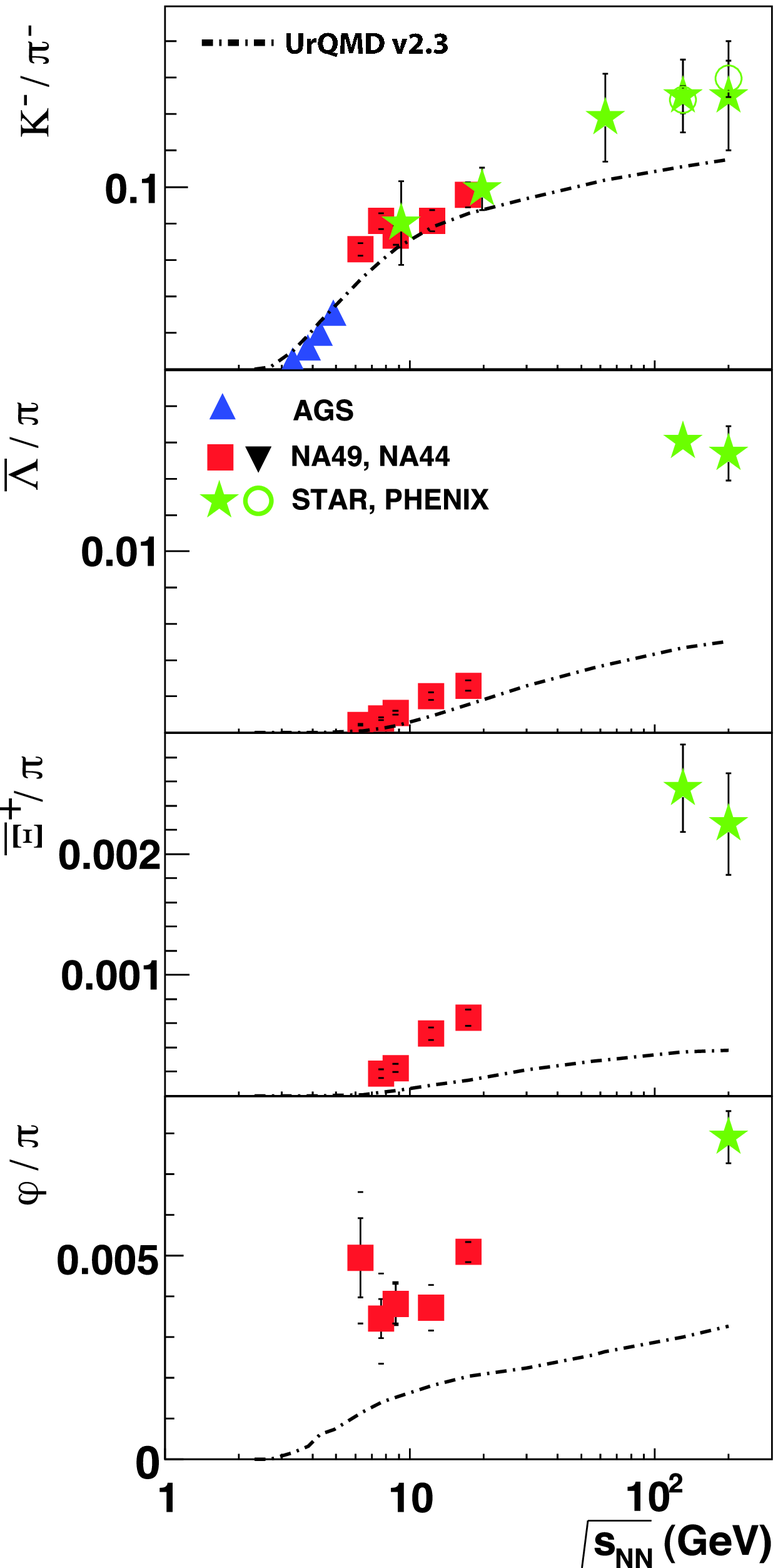}
\caption{\label{fig:fig5} The energy dependence of the midrapidity ($\mid$y$\mid$ $\leq$ 0.5) yields of strange hadrons, normalized to the pion yields ($\pi$ = 1.5 ($\pi^{-}$ + $\pi^{+}$)), in central Pb+Pb/Au+Au collisions. The data is compared to the string hadronic model UrQMD v2.3~\cite{Petersen:2008kb}.}
\end{center}
\end{figure}

Figure~\ref{fig:fig5} shows the energy dependence of relative strange particle yields at midrapidity ($\mid$y$\mid$ $\leq$ 0.5). The $K^{-}$/$\pi^{-}$, $\bar{\Lambda}$/$\pi$, $\bar{\Xi}^{+}$/$\pi$, $\Omega^{-}$+$\bar{\Omega}^{+}$/$\pi$ and $\phi$/$\pi$ ratios rise continuously from AGS to RHIC energies. In contrast the $K^{+}$/$\pi^{+}$, $\Lambda$/$\pi$ and $\Xi^{-}$/$\pi$ ratio shows a distinct maximum in the energy dependence. The results are compared to the transport model UrQMD v2.3~\cite{Petersen:2008kb}. UrQMD provides a good description for the $K^{-}$/$\pi^{-}$ and $\Lambda$/$\pi$ ratio. For other ratios large deviations are observed. Statistical hadron gas models provide better fits to the data (see e.g.~\cite{Andronic:2008gu,Becattini:2003wp}. 

\subsection{Resonances}

\begin{figure}
\begin{center}
\includegraphics[scale=0.38]{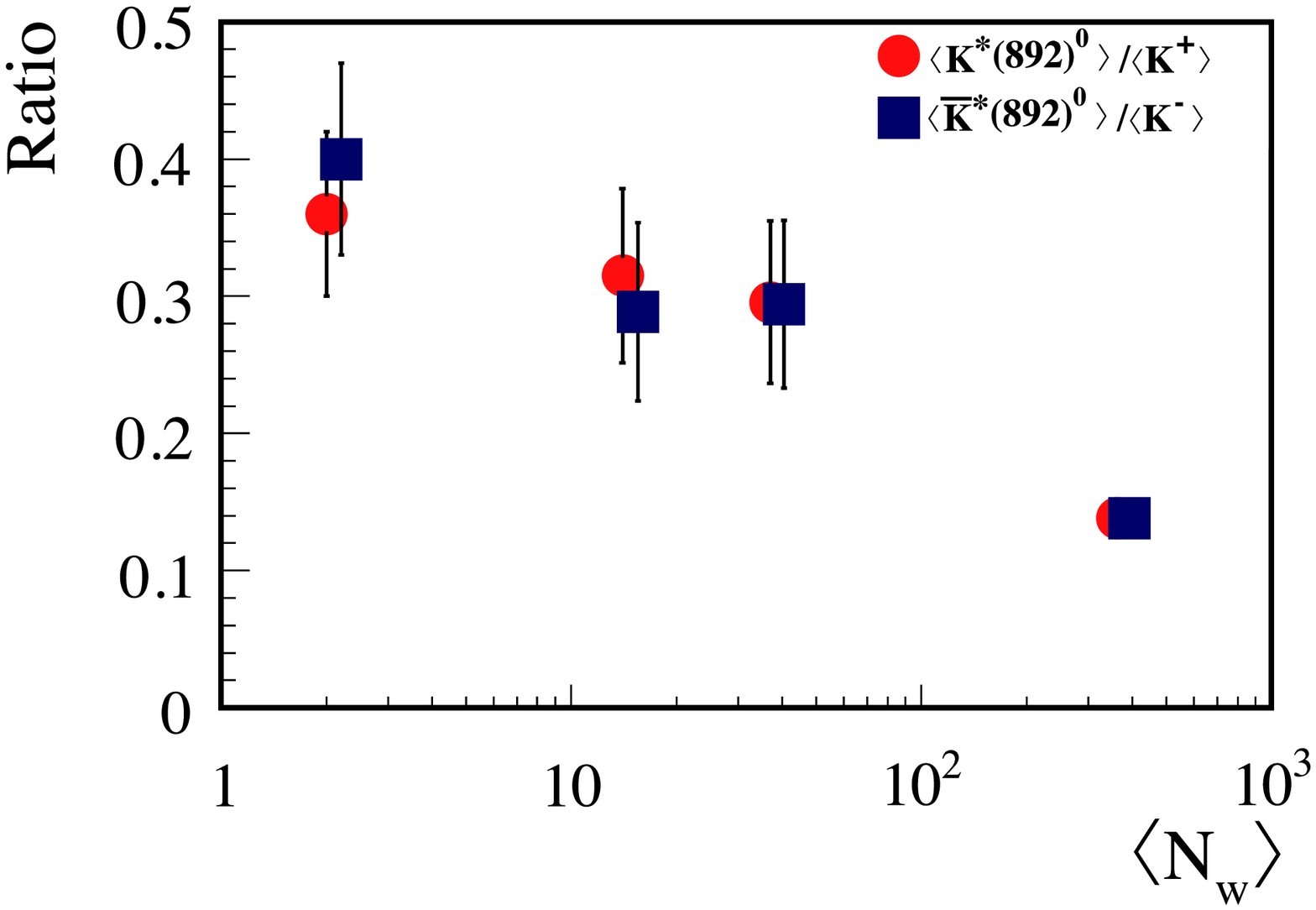}
\includegraphics[scale=0.38]{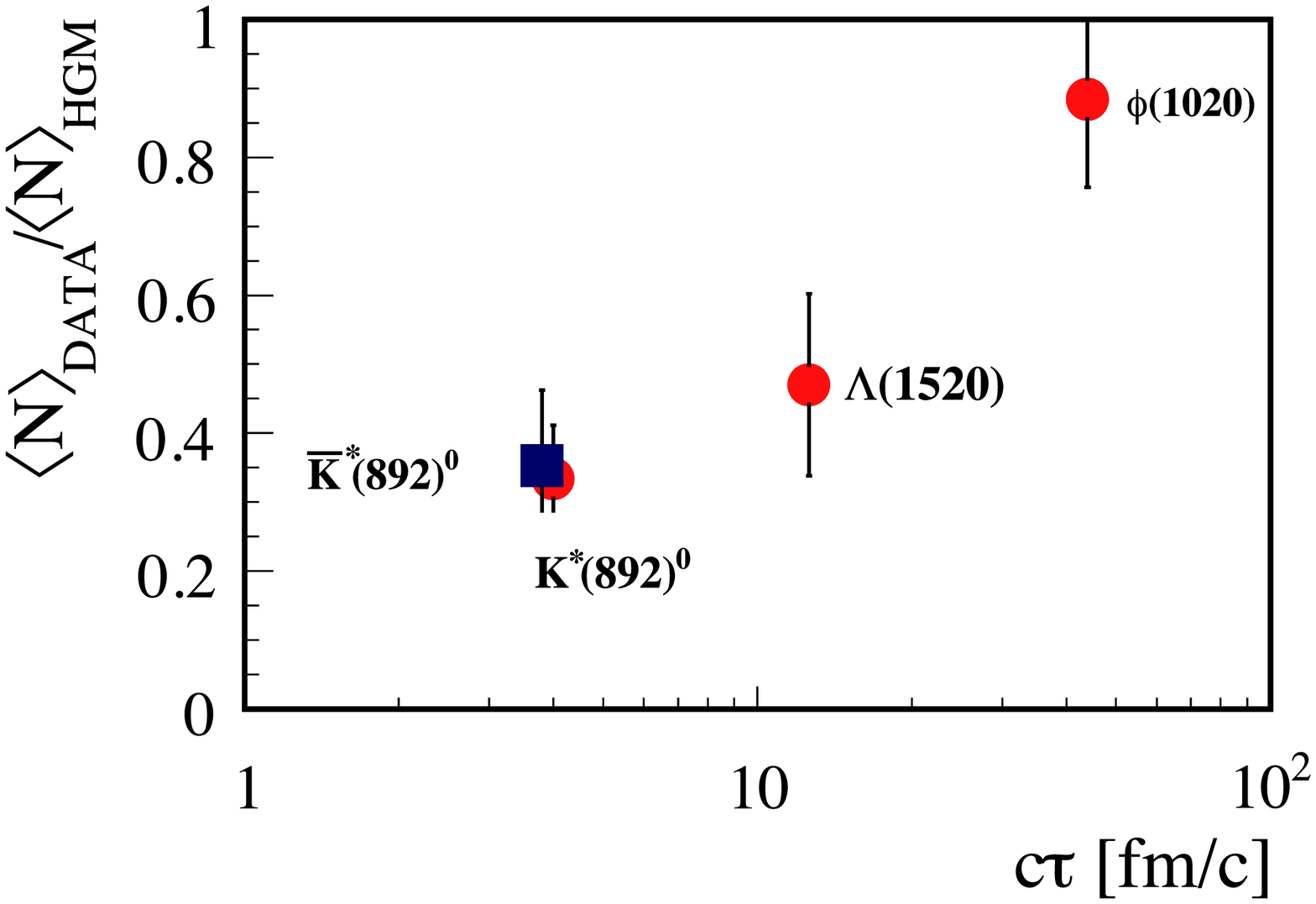}
\caption{\label{fig:fig6} Left panel: The total yield of $\left<K^{*}(892)^{0}\right>$ ($\left<\bar{K}^{*}(892)^{0}\right>$) divided by the total yields of $\left<K^{+}\right>$ ($\left<K^{-}\right>$) in p+p, C+C, Si+Si and Pb+Pb collisions at $\sqrt{s_{NN}}$ = 17.3 GeV as a function of $\left<N_{W}\right>$. Right panel: The total yield of $\left<K^{*}(892)^{0}\right>$ ($\left<\bar{K}^{*}(892)^{0}\right>$), $\Lambda(1520)$ and $\phi$ divided by a hadron gas model~\cite{Becattini:2005xt} as a function of the lifetime $c\tau$.}
\end{center}
\end{figure}

Short lived resonances are an efficient tool to probe the properties of the hot and dense medium which is created in heavy ion collisions. Resonances can be sensitive to two effects in a dense medium. If they decay before the kinetic freeze-out they may not be reconstructed due to the rescattering of the daughter particles and therefore their yield is reduced due to destruction between chemical and kinetic freeze-out. The other possibility is that after chemical freeze-out pseudo-elastic interactions~\cite{Bleicher:2002dm} among hadrons could increase the resonance population. This resonance regeneration depends on the cross-section of the interacting hadrons in the medium. Figure~\ref{fig:fig6} (left) shows the $\left<K^{*}(892)^{0}\right>$/$\left<K^{+}\right>$ and the $\left<\bar{K}^{*}(892)^{0}\right>$/$\left<K^{-}\right>$ ratios as a function of the system size at $\sqrt{s_{NN}}$ = 17.3 GeV. It is clearly visible that these ratios decrease with the increasing system size. The right panel of Fig.~\ref{fig:fig6} shows the comparison of the total yield of the resonances $\left<K^{*}(892)^{0}\right>$ ($\left<\bar{K}^{*}(892)^{0}\right>$), $\Lambda(1520)$ and $\phi$ to predictions from a hadron gas model~\cite{Becattini:2005xt}. The deviation of the hadron gas model from the data becomes larger with decreasing lifetime of the resonance. This leads to the conclusion that a large part of the reduction of the $\left<K^{*}(892)^{0}\right>$ yields could be due to rescattering of the decay daughters during the hadronic stage of the fireball and indicates that this stage lasts for a time comparable to the lifetime of the resonance. These results are comparable to published results from the STAR experiment~\cite{Adams:2004ep,Fachini:2008zz}.

\subsection{Fluctuations}

Event-by-event hadron yield ratios characterize the chemical composition of the fireball in each event and fluctuations of net baryon number and strangeness are sensitive to the properties of the early stage. Compared to other conserved quantities, like charge~\cite{Zaranek:2001di}, these fluctuations are less strongly affected by hadronic reinteractions in the later stage of the collisions. 

\begin{figure}
\begin{center}
\includegraphics[scale=0.36]{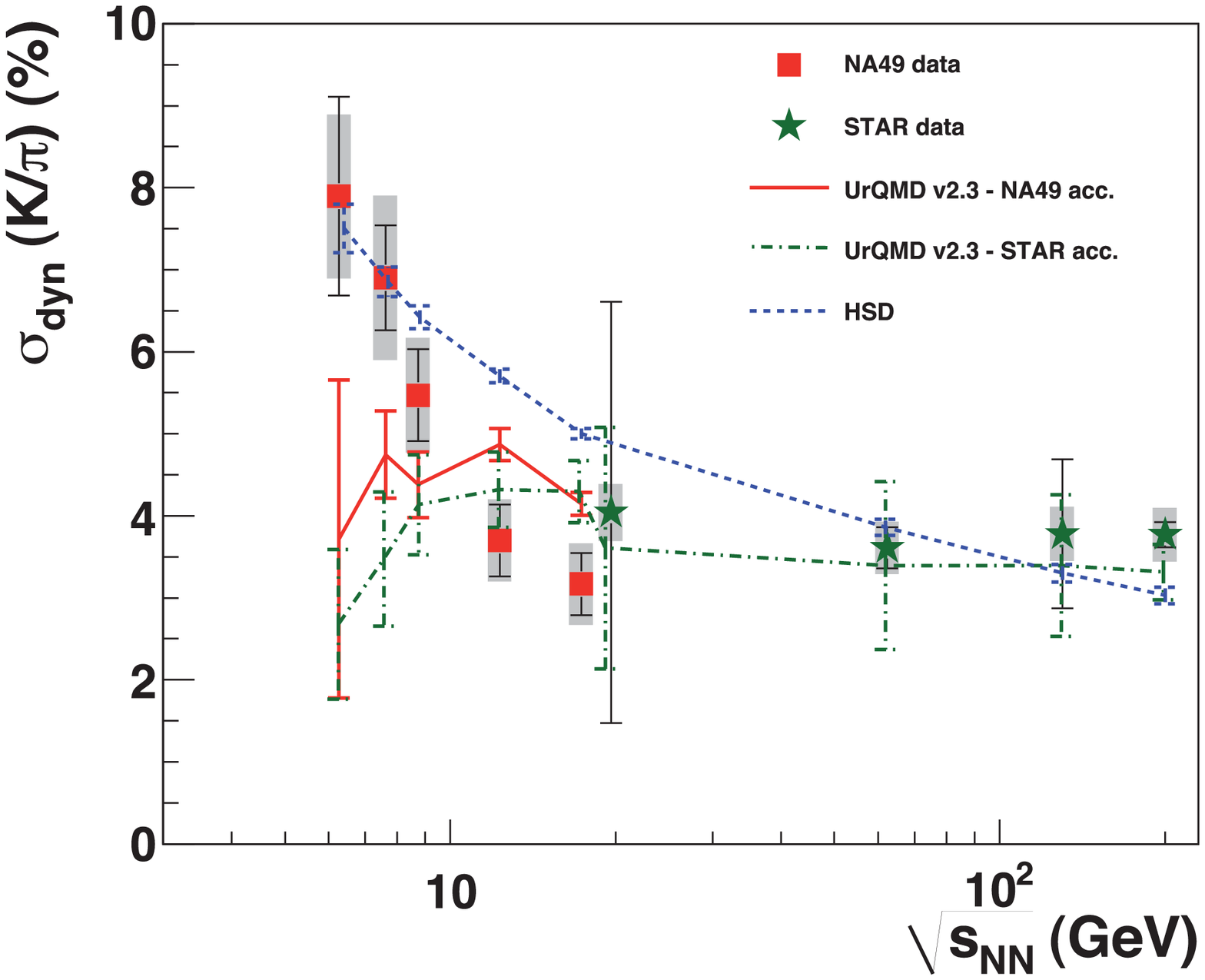}
\includegraphics[scale=0.36]{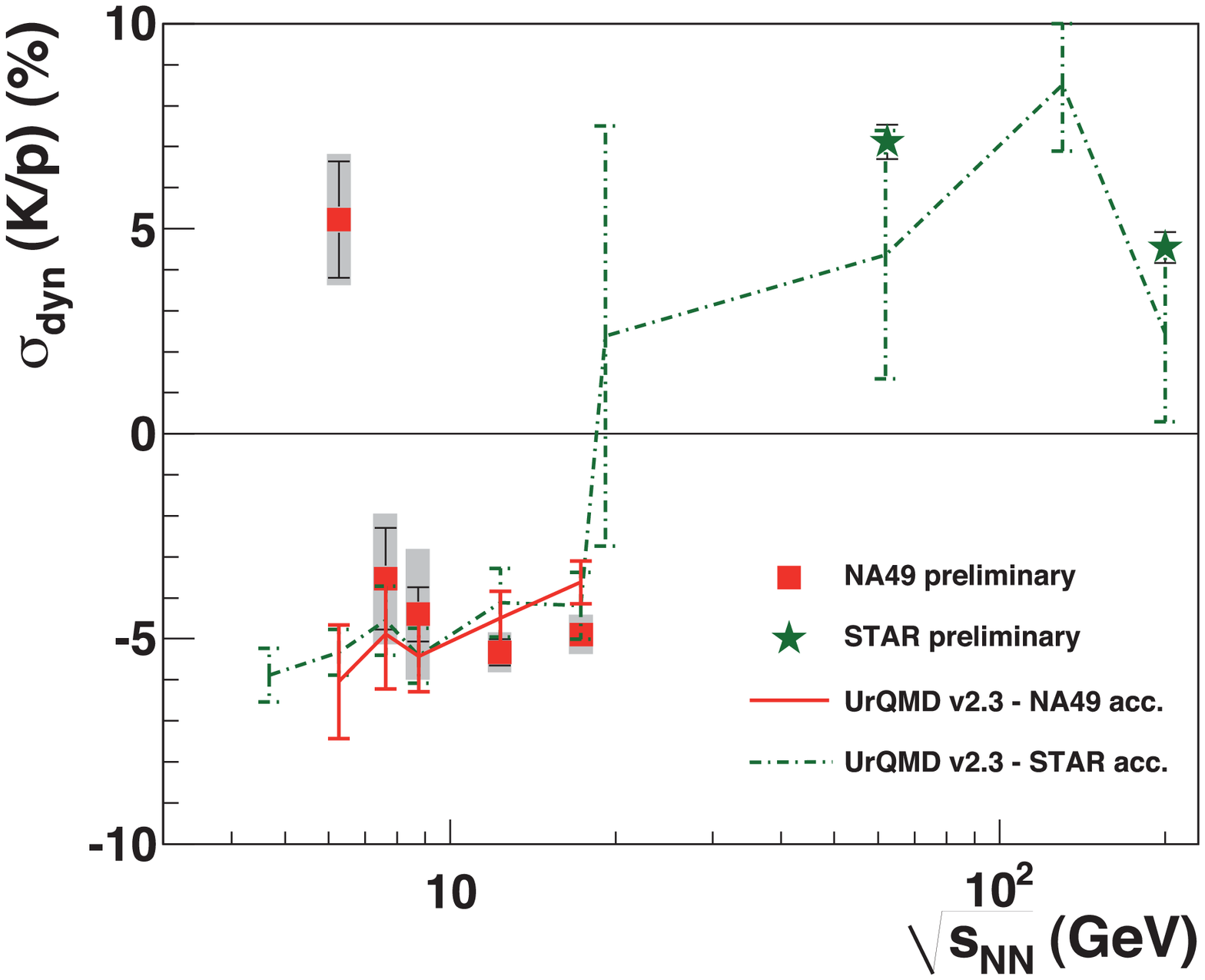}
\caption{\label{fig:fig7} Energy dependence of $\sigma_{dyn}$ for the $K/\pi$ (left panel) and $K/p$ (right panel) ratio~\cite{:2008ca,:2009if,Schuster:2009ak}. The measurements are compared to the transport models UrQMD~\cite{Schuster:2009ak} and HSD~\cite{Konchakovski:2009at}.}
\end{center}
\end{figure}

This means that possible signatures of a phase transition are less washed out. The left panel of Fig.~\ref{fig:fig7} shows the energy dependence of $\sigma_{dyn}$ for the $K/\pi$ ratio~\cite{:2008ca,:2009if,Schuster:2009ak}. A positive value for $\sigma_{dyn}$ for the $K/\pi$ ratio is observed for the whole energy range. From top SPS to top RHIC energies $\sigma_{dyn}$ is constant. The rise at low SPS energies could be a signal for the onset of deconfinement~\cite{:2008ca}. When comparing measurements to the predictions of the transport models UrQMD~\cite{Schuster:2009ak} and HSD~\cite{Konchakovski:2009at} one finds that UrQMD does not describe the increase towards lower SPS energies, whereas HSD provides a better description of the increasing trend but does not describe the data at higher SPS energies. Due to large differences between the transport model predictions no conclusions can be drawn. Furthermore it is visible that applying different acceptances (in this example NA49 and STAR acceptance) does not change the predictions for $\sigma_{dyn}$. The value of $\sigma_{dyn}$ for the $K/p$ ratio is shown in Fig.~\ref{fig:fig7} (right panel). Two sign changes are seen for this observable when looking at the energy dependence. Between $\sqrt{s_{NN}}$ = 7.6 - 17.3 a negative value is observed, whereas at $\sqrt{s_{NN}}$ = 6.3, 62.4 and 200 GeV it is positive. Comparing the results to UrQMD one finds that the model describes the overall trend except the lowest energy NA49 point.

\section*{Acknowledgments} 

This work was supported by the US Department of Energy Grant DE-FG03-97ER41020/A000, the Bundesministerium fur Bildung und Forschung, Germany, the Virtual Institute VI-146 of Helmholtz Gemeinschaft, Germany, DESY, Germany, the Polish Ministry of Science and Higher Education (1 P03B 006 30, 1 P03B 127 30, 0297/B/H03/2007/33, NN202 078735), the Hungarian Scientific Research Foundation (T032648, T032293, T043514), the Hungarian National Science Foundation, OTKA, (F034707), the Bulgarian National Science Fund (Ph-09/05), the Croatian Ministry of Science, Education and Sport (Project 098-0982887-2878) and Stichting FOM, the Netherlands.


\section*{References}
\begin{thebibliography}{40}

\bibitem{Bass:1998vz}
  S.~A.~Bass, M.~Gyulassy, H.~Stoecker and W.~Greiner,
  J.\ Phys.\ G {\bf 25}, R1 (1999)
  [arXiv:hep-ph/9810281].

\bibitem{Wang:1996yf}
  X.~N.~Wang,
  Phys.\ Rept.\  {\bf 280}, 287 (1997)
  [arXiv:hep-ph/9605214].

\bibitem{Harris:1996zx}
  J.~W.~Harris and B.~Muller,
  Ann.\ Rev.\ Nucl.\ Part.\ Sci.\  {\bf 46}, 71 (1996)
  [arXiv:hep-ph/9602235].
  
\bibitem{Rafelski:1982pu}
  J.~Rafelski and B.~Muller,
  Phys.\ Rev.\ Lett.\  {\bf 48}, 1066 (1982)
  [Erratum-ibid.\  {\bf 56}, 2334 (1986)].

\bibitem{Koch:1986ud}
  P.~Koch, B.~Muller and J.~Rafelski,
  Phys.\ Rept.\  {\bf 142}, 167 (1986).
  
\bibitem{Afanasev:1999iu}
  S.~Afanasev {\it et al.}  [NA49 Collaboration],
  Nucl.\ Instrum.\ Meth.\  A {\bf 430}, 210 (1999).

\bibitem{Anticic:2009ie}
  T.~Anticic {\it et al.}  [NA49 Collaboration],
  Phys.\ Rev.\  C {\bf 80}, 034906 (2009)
  [arXiv:0906.0469 [nucl-ex]].

\bibitem{Glauber:1970jm}
  R.~J.~Glauber and G.~Matthiae,
  Nucl.\ Phys.\  B {\bf 21}, 135 (1970).
  
\bibitem{Alt:2005zq}
  C.~Alt {\it et al.}  [NA49 Collaboration],
  Eur.\ Phys.\ J.\  C {\bf 45}, 343 (2006)
  [arXiv:hep-ex/0510009].

 \bibitem{Alt:2004wc}
  C.~Alt {\it et al.}  [NA49 Collaboration],
  Phys.\ Rev.\ Lett.\  {\bf 94}, 052301 (2005)
  [arXiv:nucl-ex/0406031].

\bibitem{Afanasiev:2002mx}
  S.~V.~Afanasiev {\it et al.}  [The NA49 Collaboration],
  Phys.\ Rev.\  C {\bf 66}, 054902 (2002)
  [arXiv:nucl-ex/0205002].
   
\bibitem{Alt:2008iv}
  C.~Alt {\it et al.}  [NA49 collaboration],
  Phys.\ Rev.\  C {\bf 78}, 044907 (2008)
  [arXiv:0806.1937 [nucl-ex]].

\bibitem{Alt:2008qm}
  C.~Alt {\it et al.}  [NA49 Collaboration],
  Phys.\ Rev.\  C {\bf 78}, 034918 (2008)
  [arXiv:0804.3770 [nucl-ex]].

\bibitem{Alt:2004kq}
  C.~Alt {\it et al.}  [NA49 Collaboration],
  Phys.\ Rev.\ Lett.\  {\bf 94}, 192301 (2005)
  [arXiv:nucl-ex/0409004].

\bibitem{Werner:2007bf}
  K.~Werner,
  Phys.\ Rev.\ Lett.\  {\bf 98}, 152301 (2007)
  [arXiv:0704.1270 [nucl-th]].
    
\bibitem{Aichelin:2008mi}
  J.~Aichelin and K.~Werner,
  Phys.\ Rev.\  C {\bf 79}, 064907 (2009)
  [arXiv:0810.4465 [nucl-th]].
    
\bibitem{Becattini:2008yn}
  F.~Becattini and J.~Manninen,
  J.\ Phys.\ G {\bf 35}, 104013 (2008)
  [arXiv:0805.0098 [nucl-th]].

\bibitem{Becattini:2008ya}
  F.~Becattini and J.~Manninen,
  Phys.\ Lett.\  B {\bf 673}, 19 (2009)
  [arXiv:0811.3766 [nucl-th]].
  
 \bibitem{Hohne:2005ks}
  C.~Hohne, F.~Puhlhofer and R.~Stock,
  Phys.\ Lett.\  B {\bf 640}, 96 (2006)
  [arXiv:hep-ph/0507276].

\bibitem{Petersen:2009zi}
  H.~Petersen, M.~Mitrovski, T.~Schuster and M.~Bleicher,
  Phys.\ Rev.\  C {\bf 80}, 054910 (2009)
  [arXiv:0903.0396 [hep-ph]].

\bibitem{Petersen:2008kb}
  H.~Petersen, M.~Bleicher, S.~A.~Bass and H.~Stocker,
  arXiv:0805.0567 [hep-ph].
  
\bibitem{Antinori:2006ij}
  F.~Antinori {\it et al.}  [NA57 Collaboration],
  J.\ Phys.\ G {\bf 32}, 427 (2006)
  [arXiv:nucl-ex/0601021].

\bibitem{:2008ez}
  B.~I.~Abelev {\it et al.}  [STAR Collaboration],
  Phys.\ Rev.\  C {\bf 79}, 034909 (2009)
  [arXiv:0808.2041 [nucl-ex]].

\bibitem{Abelev:2008zk}
  B.~I.~Abelev {\it et al.}  [STAR Collaboration],
  Phys.\ Lett.\  B {\bf 673}, 183 (2009)
  [arXiv:0810.4979 [nucl-ex]].

\bibitem{Abelev:2007xp}
  B.~I.~Abelev {\it et al.}  [STAR Collaboration],
  Phys.\ Rev.\  C {\bf 77}, 044908 (2008)
  [arXiv:0705.2511 [nucl-ex]].

\bibitem{Weber:2002qb}
  H.~Weber, E.~L.~Bratkovskaya and H.~Stoecker,
  Phys.\ Rev.\  C {\bf 66}, 054903 (2002).

\bibitem{Gazdzicki:2000ht}
  M.~Gazdzicki, M.~I.~Gorenstein and D.~Roehrich,
  arXiv:hep-ph/0006236.

\bibitem{Andronic:2008gu}
  A.~Andronic, P.~Braun-Munzinger and J.~Stachel,
  [arXiv:0812.1186 [nucl-th]].

\bibitem{Becattini:2003wp}
  F.~Becattini, M.~Gazdzicki, A.~Keranen, J.~Manninen and R.~Stock,
  Phys.\ Rev.\  C {\bf 69}, 024905 (2004)
  [arXiv:hep-ph/0310049].

\bibitem{Bleicher:2002dm}
  M.~Bleicher and J.~Aichelin,
  Phys.\ Lett.\  B {\bf 530}, 81 (2002)
  [arXiv:hep-ph/0201123].

\bibitem{Becattini:2005xt}
  F.~Becattini, J.~Manninen and M.~Gazdzicki,
  Phys.\ Rev.\  C {\bf 73}, 044905 (2006)
  [arXiv:hep-ph/0511092].

\bibitem{Adams:2004ep}
  J.~Adams {\it et al.}  [STAR Collaboration],
  Phys.\ Rev.\  C {\bf 71}, 064902 (2005)
  [arXiv:nucl-ex/0412019].

\bibitem{Fachini:2008zz}
  P.~Fachini,
  J.\ Phys.\ G {\bf 35}, 044032 (2008).

\bibitem{Zaranek:2001di}
  J.~Zaranek,
  Phys.\ Rev.\  C {\bf 66}, 024905 (2002)
  [arXiv:hep-ph/0111228].

\bibitem{:2008ca}
  C.~Alt {\it et al.}  [NA49 Collaboration],
  Phys.\ Rev.\  C {\bf 79}, 044910 (2009)
  [arXiv:0808.1237 [nucl-ex]].

\bibitem{:2009if}
  B.~I.~Abelev {\it et al.}  [STAR Collaboration],
  Phys.\ Rev.\ Lett.\  {\bf 103}, 092301 (2009)
  [arXiv:0901.1795 [nucl-ex]].

\bibitem{Schuster:2009ak}
  T.~Schuster  [for the NA49 collaboration],
  PoS C {\bf POD2009}, 029 (2009)
  [arXiv:0910.0558 [nucl-ex]].

\bibitem{Konchakovski:2009at}
  V.~P.~Konchakovski, M.~Hauer, M.~I.~Gorenstein and E.~L.~Bratkovskaya,
  J.\ Phys.\ G {\bf 36}, 125106 (2009)
  [arXiv:0906.3229 [nucl-th]].

\end{thebibliography}
\end{document}